\pdfoutput=1
\documentclass[aps,prb,preprint,floatfix,groupedaddress,amsmath,amssymb,amsfonts]{revtex4-1}
\usepackage{mathptmx}                    
\usepackage[Symbol]{upgreek}
\usepackage[utf8]{inputenc}
\usepackage{graphicx}
\usepackage[dvipsnames]{xcolor}
\usepackage{textcomp}

\usepackage[
,textwidth=17.5cm
,textheight=23.5cm
,verbose
,pdftex
]{geometry}
\usepackage{xspace}

\usepackage[pdftex]{hyperref}
\hypersetup{
  pdftitle={Luminescent N-polar (In,Ga)N/GaN quantum wells grown by plasma-assisted molecular beam epitaxy at high temperature},
	colorlinks,
	citecolor=blue
	}
\usepackage{cleveref}

\newcommand{\celsius}{$^{\circ}$C\xspace}
\makeatletter
\newcommand*{\refig}[2]{\hyperref[#1]{\ref*{#1}(#2)}}
\makeatother

\DeclareMathAlphabet{\mathsf}{OT1}{\sfdefault}{m}{n}
\SetMathAlphabet{\mathsf}{bold}{OT1}{\sfdefault}{b}{n}

\begin{document}

\title{Luminescent N-polar (In,Ga)N/GaN quantum wells grown by plasma-assisted molecular beam epitaxy at high temperature}

\author{C.~Ch\`{e}ze}
\email[Electronic mail: ]{cheze@pdi-berlin.de}
\author{F.~Feix}
\author{J.~L\"{a}hnemann}
\author{T.~Flissikowski}
\affiliation{Paul-Drude-Institut f\"ur Festk\"orperelektronik, Hausvogteiplatz 5--7, 10117 Berlin, Germany}
\author{M.~Kry\'{s}ko}
\author{P. Wolny}
\author{H. Turski}
\affiliation{Institute of High Pressure Physics, PAS, Soko{\l}owska 29/37, 01142 Warszawa, Poland}
\author{C.~Skierbiszewski}
\affiliation{Institute of High Pressure Physics, PAS, Soko{\l}owska 29/37, 01142 Warszawa, Poland}
\affiliation{TopGaN Ltd., Soko{\l}owska 29/37, 01142 Warszawa, Poland}
\author{O.~Brandt}
\affiliation{Paul-Drude-Institut f\"ur Festk\"orperelektronik, Hausvogteiplatz 5--7, 10117 Berlin, Germany}

\begin{abstract}
N-polar (In,Ga)N/GaN quantum wells prepared on freestanding GaN substrates by plasma-assisted molecular beam epitaxy at conventional growth temperatures of about 650\,\celsius do not exhibit any detectable luminescence even at 10\,K. In the present work, we investigate (In,Ga)N/GaN quantum wells grown on Ga- and N-polar GaN substrates at a constant temperature of 730\,\celsius. This exceptionally high temperature results in a vanishing In incorporation for the Ga-polar sample. In contrast, quantum wells with an In content of 20\% and abrupt interfaces are formed on N-polar GaN. Moreover, these quantum wells exhibit a spatially homogeneous green luminescence band up to room temperature, but the intensity of this band is observed to strongly quench with temperature. Temperature-dependent photoluminescence transients show that this thermal quenching is related to a high density of nonradiative Shockley-Read-Hall centers with large capture coefficients for electrons and holes.
\end{abstract}


\maketitle
N-polar (i.\,e., $[000\bar{1}]$-oriented) group-III nitride heterostructures are currently attracting interest as potential candidates for advanced electronic and optoelectronic devices. One of the unique advantages of these structures as compared to their well-established Ga-polar counterparts is the opposite direction of the polarization-induced internal electrostatic fields. Specifically, these reversed fields have the potential to improve the scalability of GaN high-frequency transistors by providing stronger electron confinement and reducing ohmic contact resistance,\cite{Wong2013} to enhance the performance of light emitting diodes by increasing the internal quantum efficiency and the carrier injection efficiency,\cite{Akyol2011, Feng2015, Dong2012, Han2012} and to increase the collection efficiency of the photocurrent in solar cells.\cite{Li2011}

Concerning light emission from N-polar (In,Ga)N/GaN quantum wells (QWs), however, a serious problem has recently been identified for samples synthesized by plasma-assisted molecular beam epitaxy (PAMBE). For homoepitaxial samples prepared on free-standing GaN substrates, no photoluminescence (PL) signal was detected even at 10\,K, while the intense PL band observed for heteroepitaxial samples grown on SiC was found to originate exclusively from semipolar facets around $\vee$ pits induced by threading dislocations.\cite{Cheze2013,Fernandez2016} Spatially localized emission has also been reported for samples fabricated by metal-organic chemical vapor deposition (MOCVD), in this case stemming from semipolar QWs formed around hexagonal mounds.\cite{Song2015} Since thick (In,Ga)N layers exhibited detectable PL, \citet{Cheze2013} and \citet{Fernandez2016} attributed their finding to a high concentration of nonradiative point defects at the interfaces between the QWs and the barriers.\cite{Cheze2013, Fernandez2016} 

The samples used for the investigations of \citet{Cheze2013} and \citet{Fernandez2016} were fabricated under standard conditions optimized for Ga-polar structures as also done in previous work.\cite{Akyol2011, Nath2010a} However, it has been established that N-polar (In,Ga)N layers can be grown at substantially higher temperatures compared to their Ga-polar counterparts.\cite{Yoshikawa2008, Koblmuller2007a, Nath2010a, Keller2007, Shojiki2015} The higher thermal stability of N-polar group-III nitrides is related to their specific bonding configuration, causing surface N atoms to be more strongly bound by three back bonds to the cation layer underneath compared to the single back bond for the metal-polar case.\cite{Yoshikawa2008} Moreover, for N-polar (In,Ga)N layers grown by PAMBE, it was shown that an increase of the growth temperature from $500$ to $600$\,\celsius enhances the PL intensity.\cite{Nath2010a}

In the present work, we investigate N-polar (In,Ga)N/GaN QWs synthesized at a growth temperature usually only used for bare GaN, namely, 730\,\celsius.\cite{Skierbiszewski2004} X-ray diffraction (XRD) profiles evidence the formation of highly uniform QWs with abrupt interfaces and an In content of 20\%. The sample exhibits a PL band in the green spectral range up to room temperature. Using monochromatic catholuminescence (CL) maps, we ensure that this luminescence band originates from the N-polar QWs. Time-resolved PL measurements reveal that recombination occurs between spatially separate, individually localized electrons and holes. A quantitative analysis of the transients shows that the strong thermal quenching of the PL intensity is related to Shockley-Read-Hall centers with high carrier capture rates.

\begin{figure*}
{\includegraphics*[width=12cm]{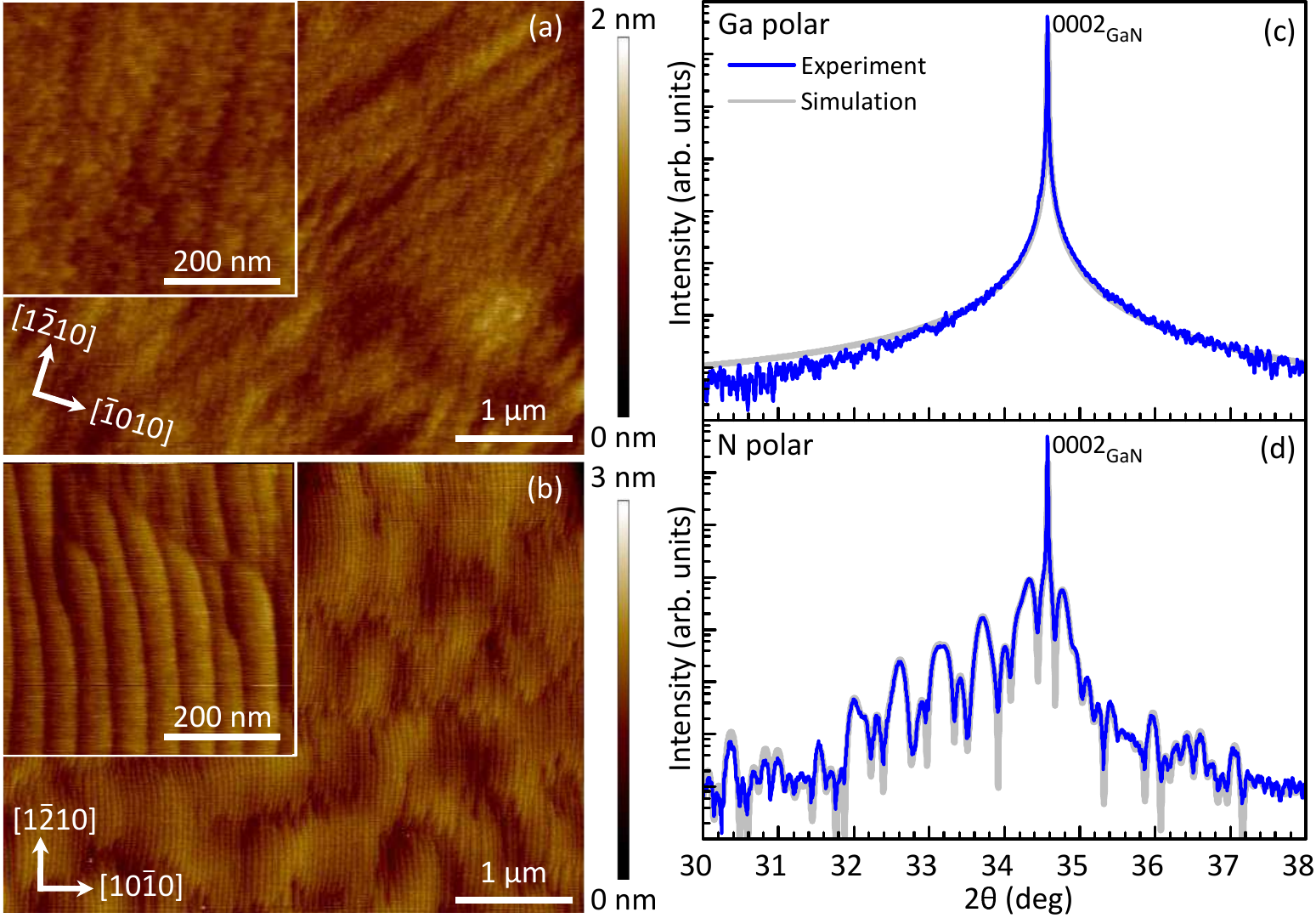}}
{\caption{Surface morphology and interface definition of the Ga- and N-polar samples. Representative AFM topographs of the (a) Ga- and (b) N-polar (In,Ga)N/GaN triple QWs under investigation. The lateral and vertical scales are indicated in the figure. Atomic steps are clearly observed in the magnified area presented in the insets. Experimental and simulated $\upomega$-$2\uptheta$ XRD profiles of the (c) Ga- and (d) N-polar samples across the 0002 reflection of GaN. The simulation in (c) shows the profile of bare GaN.}\label{Fig1}}
\end{figure*}

The two samples under investigation were grown in a Veeco GEN20 PAMBE system equipped with a radio frequency plasma source for generating active N and solid-source effusion cells for Ga and In. As substrates, we used free-standing Ga-polar GaN wafers with dislocation densities ranging from 10$^{7}$ to 10$^{8}$\,cm$^{-2}$ produced by Saint-Gobain using hydride vapor phase epitaxy. Re-polishing of the Ga-polar (0001) and N-polar (000$\bar{1}$) surfaces was carried out at the Institute of High Pressure Physics and resulted in smooth surfaces with a miscut of 0.5$^\circ$ towards a direction close to $\langle10\bar{1}0\rangle$.
The growth experiment was carried out on both substrate orientations simultaneously. The temperature at the growth front was determined from Ga desorption rates using in-situ laser reflectometry (LR) calibrated with a band-edge spectrometer BandiT from k-Space Associates.\cite{Siekacz2009, Cheze2013} The metal and N fluxes, expressed as effective growth rates, have been determined as described in Ref.~\onlinecite{Cheze2013}. Throughout the entire growth run, the temperature and N flux were kept constant at 730\,\celsius and 4.4\,$\times\,10^{14}$\,atoms\,cm$^{-2}$\,s$^{-1}$, respectively. Following the deposition of a $50$-nm-thick GaN buffer, an (In,Ga)N/GaN triple QW structure was grown. During QW growth, the Ga and In fluxes were set to 0.7\,$\times\,10^{14}$ and 7.5\,$\times\,10^{14}$\,atoms\,cm$^{-2}$\,s$^{-1}$, respectively, while a Ga flux of 7.8\,$\times\,10^{14}$\,atoms\,cm$^{-2}$\,s$^{-1}$ was used for the GaN barrier growth. Owing to the high growth temperature, the use of In as a surfactant during barrier growth\cite{Cheze2013, Fernandez2016} was rendered superfluous. Subsequently, the growth was interrupted for the desorption of excess metal, which was monitored in situ by LR. This procedure was carried out three times, and completed by the growth of a 24\,nm GaN cap layer.

The surface morphology of the structures was examined by atomic force microscopy (AFM), while their periodicity, interface abruptness, and composition were assessed by triple-crystal $\upomega$-$2\uptheta$ XRD scans performed with CuK$_{\upalpha1}$ radiation using a diffractometer equipped with a Ge(220) hybrid monochromator and a Ge(220) analyzer crystal. The emission of the samples was investigated by PL spectroscopy under both steady-state and pulsed excitation. PL spectra were obtained by exciting the samples by a He-Cd laser ($\uplambda_\text{L} = 325$\,nm) with an intensity of 10\,kW/cm$^{2}$. The emitted light was spectrally dispersed by a monochromator and detected with a liquid-nitrogen-cooled charge-coupled device. PL transients were acquired using a frequency-doubled, femtosecond Ti:sapphire laser ($\uplambda_\text{L} = 349$\,nm) for excitation with an energy fluence per pulse of 3\,\textmu J\,cm$^{-2}$ (corresponding to a maximum charge carrier density of approximately $3 \times 10^{11}$\,cm$^{-2}$) and a repetition rate of 9.3\,kHz. We employed time-correlated single-photon counting for detection and recorded the transients at the PL peak energy within a spectral range of about 20\,meV. The temporal resolution amounted to 45\,ps for the investigation of the initial (up to 1\,\textmu s) decay of the PL intensity $I_\text{PL}$, and to 500\,ps for transients recorded over the full time range of 33\,\textmu s. The CL measurements were carried out at an acceleration voltage of 5\,kV and monochromatic maps were acquired within a spectral range of about 30\,meV. For temperature-dependent experiments, the samples were mounted in He-flow cryostats.

Figures~\refig{Fig1}{a} and~\refig{Fig1}{b} display the morphology of the Ga- and N-polar samples over an area of $5 \times 5$\,\textmu m$^2$. For both samples, the root mean square roughness amounts to less than 0.2\,nm, and regular atomic steps are observed evidencing step flow growth. Obviously, the growth conditions employed are suitable for  both Ga- and N-polar surfaces as far as the morphology is concerned. 

\begin{figure*}
{\includegraphics*[width=12cm]{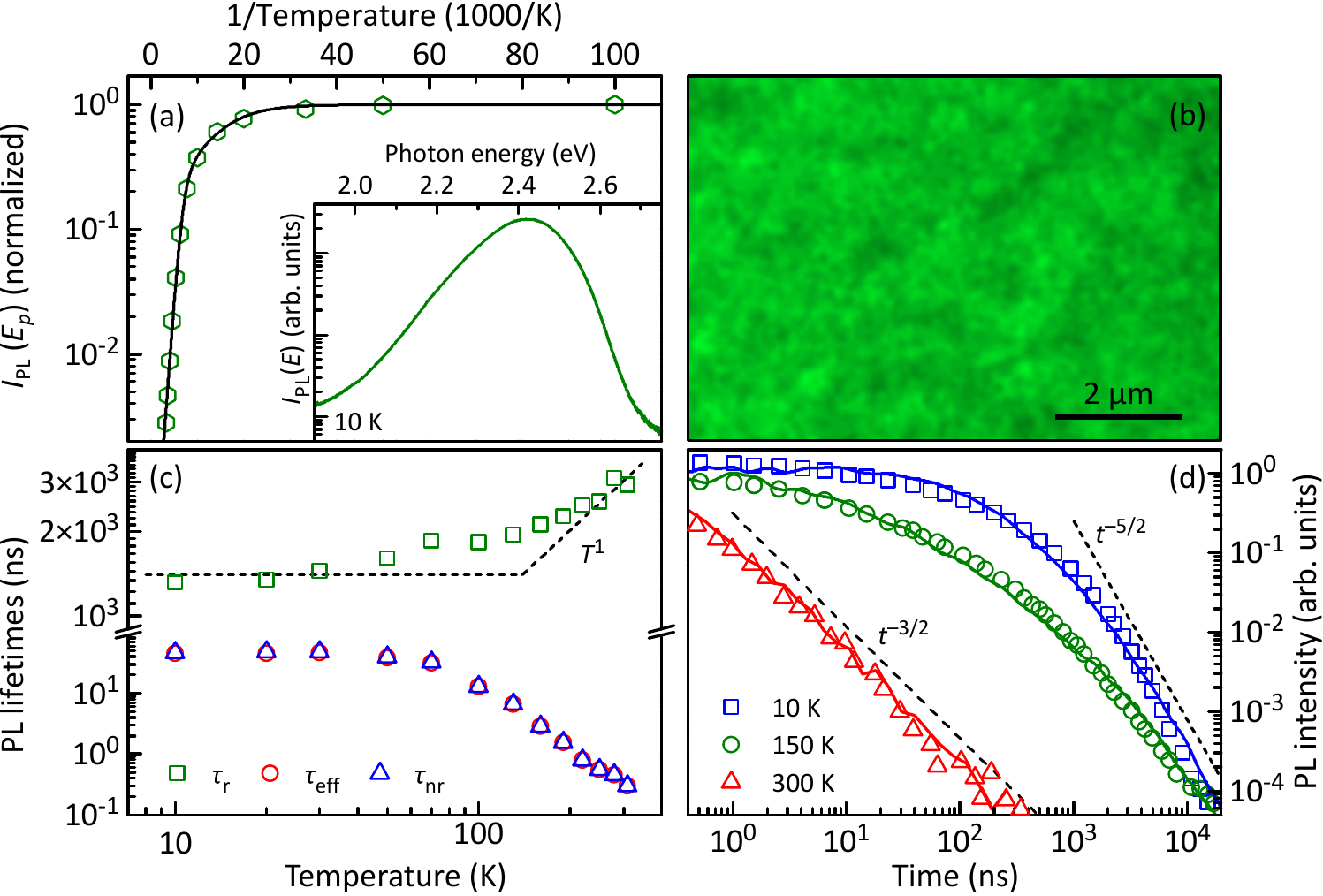}}
{\caption{Emission characteristics of the N-polar triple QW sample. (a) Normalized Arrhenius representation of the temporally integrated peak PL intensity $I_\text{PL}(E_p)$ (symbols) with a corresponding fit (line). Inset: PL spectrum recorded under steady-state conditions at 10\,K. (b) True color representation of the monochromatic CL map recorded at 2.42\,eV and 30\,K. The normalized intensity in the map ranges from 0.6 (dark) to 1 (bright). (c) Effective ($\tau_\text{eff}$), radiative ($\tau_\text{r}$), and nonradiative ($\tau_\text{nr}$) PL lifetimes as a function of temperature ($T$). (d) Double-logarithmic representation of experimental PL transients (symbols) and corresponding simulations (lines). The dashed lines indicate power laws which follow the time ($t$) dependence denoted in the figure.}\label{Fig2}}
\end{figure*}
Figures~\refig{Fig1}{c} and~\refig{Fig1}{d} show $\upomega$-$2\uptheta$ scans across the 0002 reflection of GaN for the Ga- and N-polar samples, respectively. For the Ga-polar sample, the profile lacks any feature besides the GaN 0002 reflection, and coincides with a simulated XRD profile of a bare GaN layer, revealing that no In was incorporated during growth. In contrast, the XRD profile of the N-polar sample exhibits pronounced satellite reflections, which are reproduced accurately by a simulation based on dynamical diffraction theory [see Fig.~\refig{Fig1}{d}] for three periods of $4.4 \pm 0.2$\,nm In$_{0.20}$Ga$_{0.80}$N QWs separated by $12.0 \pm 0.2$\,nm GaN barriers. These results demonstrate that the growth temperature of 730\,\celsius is too high for growth of Ga-polar (In,Ga)N, but perfectly suitable for the synthesis of N-polar (In,Ga)N/GaN QWs with abrupt heterointerfaces and high uniformity. Transmission electron microscopy on a similar sample (see Fig.~10 in Ref.~\onlinecite{Schulz2014}) further attests to the high structural perfection of the N-polar (In,Ga)N/GaN QWs, particularly regarding the atomically abrupt $(000\bar{1})$ interfaces and the uniform, random In distribution. 


We now turn to the PL properties of the N-polar (In,Ga)N/GaN QWs. The inset in Fig.~\refig{Fig2}{a} shows an exemplary PL spectrum of this sample obtained at 10\,K under steady-state conditions. The emission band from the QWs peaks at around 2.42\,eV and has a full width at half maximum of about 220\,meV. This comparatively large width indicates the presence of disorder inevitable in the random alloy (In,Ga)N,\cite{Schulz2015} but partly also results from the overlap with the yellow luminescence band of the GaN substrate,\cite{Xu2012, Reshchikov2005} as revealed by CL spectra acquired at different acceleration voltages and thus excitation depths (not shown here). The temperature dependence of the PL peak intensity shown in Fig.~\refig{Fig2}{a} is obtained by temporally integrating PL transients measured after pulsed excitation. At temperatures above 70\,K, a strong thermally activated quenching of the PL intensity sets in. The activation energy, deduced from a fit with the standard three-level model,\cite{Bimberg1971} amounts to $100 \pm 10$\,meV. Due to this high activation energy, the PL intensity quenches by a factor of 350 between 10 and 300\,K.

The monochromatic CL map displayed in Fig.~\refig{Fig2}{b} demonstrates that the emission does not originate from isolated hillocks or pits as observed previously for N-polar (In,Ga)N/GaN QWs with high dislocation density,\cite{Song2015,Fernandez2016} but is spatially homogeneous. The emission is thus characteristic for (In,Ga)N/GaN QWs bound by polar $\{0001\}$ planes, and its strong quenching is likely to be caused by the same interfacial point defects that have been made responsible for the absence of any detectable emission from N-polar (In,Ga)N/GaN QWs grown at lower substrate temperatures.\cite{Cheze2013,Fernandez2016} 

In order to quantitatively understand the recombination mechanisms governing the light emission of the N-polar sample, we perform and analyze time-resolved PL experiments. First, we deduce temperature-dependent PL lifetimes from the initial decay of PL transients recorded with high temporal resolution. To this end, we define a phenomenological effective PL lifetime ($\tau_\text{eff}$) as the time at which the peak PL intensity has decreased to $1/e$ of its initial value. Additionally, we suppose that $1/\tau_\text{eff} = 1/\tau_\text{r} + 1/\tau_\text{nr}$ with the radiative and nonradiative lifetimes $\tau_\text{r}$ and $\tau_\text{nr}$, respectively. The temperature dependence of $\tau_\text{r}$ is given by the inverse PL peak intensity of the transient just after the laser pulse.\cite{Brandt2002} To obtain absolute values for $\tau_\text{r}$ and $\tau_\text{nr}$, we compare the temporally integrated PL intensity at 10\,K to that of a Ga-polar reference QW emitting in the blue spectral range\cite{Feix2017} mounted side-by-side. Since we know that the internal quantum efficiency ($\eta = \tau_\text{eff} / \tau_\text{r}$) of this reference sample is close to 1 at 10\,K,\cite{Feix2017} the comparison of the integrated intensities yields $\eta \approx 0.03$ for the N-polar sample at 10\,K. Consequently, $\tau_\text{r} \gg \tau_\text{nr}$ and $\tau_\text{eff} \approx \tau_\text{nr}$ even at low temperatures as seen in Fig.~\refig{Fig2}{c}. 

The temperature dependence of $\tau_\text{r}$ shown in Fig.~\refig{Fig2}{c} is exactly the same as the one observed for Ga-polar (In,Ga)N/GaN QWs: it is almost constant up to 50\,K, reflecting that recombination occurs from localized states, before it gradually approaches the linear dependence expected for a two-dimensional system (such as a QW) at higher temperatures.\cite{Waltereit2001} However, at temperatures exceeding 70\,K, $\tau_\text{nr}$ decreases steeply, dropping from a value of 30\,ns at 70\,K to 0.3\,ns at 300\,K. It is this decrease that is mainly responsible for the quenching of the PL intensity in the same temperature interval.  

\begin{table}[b]
 \caption{Parameters for the simulation of the temperature-dependent PL transients. Recombination is controlled by the tunneling range $a$, the diffusivity $D$, and the radiative recombination and nonradiative capture coefficients $B_0$ and $b_0$, respectively.}
 \label{Tab1}
 \begin{ruledtabular}
\begin{tabular}{c c c c}
parameter & 10\,K & 150\,K & 300\,K\\
\hline
$a$ (nm)&  6 & 4 & 5 \\
$D$ (nm$^2$/ns) & 0 & 0 & 0.11\\
$B_0$ (ns$^{-1}$) & $3.0 \times 10^{-4}$ & $5.5 \times 10^{-4}$ & 0\\
$b_0$ (ns$^{-1}$) & 0.025 & 1 & 50\\
\end{tabular}
\end{ruledtabular}
\end{table}


To gain further insight into the recombination dynamics, we examine the actual time dependence of the PL decay in detail. Figure~\refig{Fig2}{d} shows PL transients recorded at three different temperatures on a double-logarithmic scale. Clearly, the decay asymptotically obeys a power law independent of temperature, revealing that recombination occurs between spatially separated, individually localized electrons and holes.\cite{Feix2017} In this case, recombination can occur by tunneling and/or diffusion (with a subsequent radiative annihilation of electrons and holes at the same spatial location). With increasing temperature, the decay accelerates (but still follows a power law) and the temporally integrated PL intensity decreases strongly, reflecting the presence of nonradiative recombination. Utilizing the reaction-diffusion equations introduced in Refs.~\onlinecite{Sabelfeld2015,Feix2017} and the parameters listed in Table~\ref{Tab1}, we obtain simulated transients matching the experimental ones. For all simulations, we assumed a density of nonradiative centers of $1.9 \times 10^{-3}$\,nm$^{-2}$ ($1.9 \times 10^{11}$\,cm$^{-2}$), which is only six times higher than that of the Ga-polar reference. 

In fact, the reason for the low quantum efficiency of the N-polar sample as compared to the Ga-polar reference is not an especially high density of nonradiative centers, but the comparatively small radiative ($B_0$) and large nonradiative ($b_0$) coefficients. This scenario is particularly pronounced at room temperature, where we observe an exact $t^{-3/2}$ dependence indicating a diffusion-controlled recombination process. Indeed, the transient can be reproduced by allowing for a finite diffusivity and setting radiative tunneling recombination to zero (cf.\ Table~\ref{Tab1}). Simultaneously, $b_0$ has to be increased significantly to account for the loss in PL intensity with increasing temperature, consistent with the results obtained from the phenomenological analysis summarized in Fig.~\refig{Fig2}{c}. 

The smaller radiative coefficients are easily understood as being a consequence of the comparatively large well width of the present sample, the larger internal fields due to the high In content, and the reduction of the momentum matrix element with increasing In content due to stronger localization effects.\cite{AufDerMaur2016} The larger nonradiative coefficients seem to suggest a different nature of the nonradiative centers in Ga- and N-polar (In,Ga)N/GaN QWs.

To check for a possible extrinsic origin of the nonradiative centers in the N-polar QWs, we have investigated two N-polar QW structures grown at 660 and 730\,\celsius by secondary ion mass spectrometry (SIMS). As reported before in Ref.~\onlinecite{Cheze2013}, the sample grown at 660\,\celsius does not emit any detectable PL or CL, in contrast to the one fabricated at 730\,\celsius, which has similar optical properties as the sample investigated in the present work. 
 
Four elements were found in notable concentrations in these two samples: B, C, O, and Ca. Specifically, the SIMS profiles (not shown here) imply that the B, C, O, and Ca concentrations were reduced from 2 to $1 \times 10^{17}$\,cm$^{-3}$, from 6 to $2 \times 10^{17}$\,cm$^{-3}$, from 20 to $4 \times 10^{17}$\,cm$^{-3}$, and from 5 to $3 \times 10^{17}$\,cm$^{-3}$, respectively, by increasing the growth temperature from 660 to 730\,\celsius. At these low concentrations, B is an isoelectronic impurity specific to the PAMBE growth process and is found with virtually identical concentrations in Ga-polar samples. The C and O concentrations are still about an order of magnitude higher than those observed in Ga-polar samples prepared in the same MBE system or in N-polar (In,Ga)N/GaN QWs grown by metal-organic chemical vapor deposition.\cite{Keller2014, Lund2017} However, at these modest levels, both C and O should predominantly incorporate substitutionally on N sites, and are not expected to induce nonradiative processes unless they form complexes with native defects.\cite{Dreyer2016}

In contrast, Ca was recently identified to act as a deep nonradiative Shockley-Read-Hall center in GaN. Moreover, it was found to incorporate with a concentration that strongly increases with decreasing growth temperature, resulting in concentrations up to $10^{18}$\,cm$^{-3}$ in Ga-polar (In,Ga)N/GaN QWs grown at 600\,\celsius.\cite{Young2016a,Shen2017} For the present sample, the experimentally measured volume density of Ca corresponds to $1.3 \times 10^{11}$\,cm$^{-2}$ within one QW, close to the value of $1.9 \times 10^{11}$\,cm$^{-2}$ assumed for the nonradiative centers in the simulated PL transients in Fig.~\refig{Fig2}{d}.  
     
Even more intriguing is the fact that the electron capture coefficient of Ca is theoretically predicted to increase by several orders of magnitude when changing the band gap of (In,Ga)N from 2.7 to 2.2\,eV.\cite{Shen2017} In other words, a certain concentration of Ca may go unnoticed for a blue emitting (In,Ga)N/GaN QW, but may have a devastating effect on the internal quantum efficiency of a green emitting one. This peculiar behavior is consistent with the results of our time-resolved measurements, which indicated a similar density of nonradiative centers for blue emitting Ga- and green emitting N-polar samples, but much larger capture coefficients for the latter ones. 
 
In conclusion, we have shown that the use of an exceptionally high growth temperature in PAMBE facilitates the synthesis of N-polar (In,Ga)N/GaN QWs emitting in the green spectral range. Still, the luminous efficiency of these samples is not comparable with the one of Ga-polar (In,Ga)N/GaN QWs, but is limited by a highly efficient nonradiative recombination channel attributed to a high density of interfacial point defects. The impurity Ca seems to be a potential candidate for this as yet unknown point defect. Future work toward the identification of this defect should investigate ways to reduce or avoid the incorporation of Ca, such as growth at even higher temperatures, suitable substrate cleaning procedures, and inclusion of buried layers to getter Ca at the heterointerfaces,\cite{Young2016a} and should correlate the residual Ca content with the luminous efficiency of the QWs.


Funding of this work by the European Union's FP7-PEOPLE-IAPP-2008 programme under grant agreement No.\ 230765 (SINOPLE) and FP7-NMP-2013-SMALL-7 programme under grant agreement No.\ 604416 (DEEPEN), and partially by the Polish National Centre for Research and Development Grant PBS3/A3/23/2015 is gratefully acknowledged. We thank S. Fern\'andez-Garrido for a critical reading of the manuscript, M.~Bo\'{c}kowski, A. Feduniewicz-Zmuda and B. Grzywacz for the substrate preparation.

\end{document}